\documentclass[preprint, 5p, number, sort&compress, lefttitle]{elsarticle}

\usepackage{amsmath}
\usepackage{amssymb}
\usepackage{graphicx}
\usepackage{bm}
\usepackage{hyperref}
\hypersetup{
  colorlinks=true,
  linkcolor=blue,
  citecolor=blue,
  urlcolor=blue,
}

\def\fund{National Natural Science Foundation of China~(Grant No.~12075003)}

\def\declaration{The authors declare that they have no known competing financial interests or personal relationships that could have appeared to influence the work reported in this paper.}

\usepackage{ulem}
\usepackage{xcolor}

\journal{Astroparticle Physics 148 (2023) 102831 \href {https://doi.org/10.1016/j.astropartphys.2023.102831}{doi:10.1016/j.astropartphys.2023.102831}}

\begin{document}


\begin{frontmatter}

\title{Lorentz invariance violation induced threshold anomaly versus very-high energy cosmic photon emission from GRB 221009A}

\author[aff1]{Hao Li}
\ead{haolee@pku.edu.cn}

\author[aff1,aff2,aff3]{Bo-Qiang Ma\corref{cor1}}
\ead{mabq@pku.edu.cn}
\cortext[cor1]{Corresponding author}

\affiliation[aff1]{
    organization = {School of Physics,
    Peking University},
    city = {Beijing 100871},
    country = {China}}

\affiliation[aff2]{
    organization = {Center for High Energy Physics, Peking University},
    city = {Beijing 100871},
    country = {China}}

\affiliation[aff3]{
    organization = {Collaborative Innovation Center of Quantum Matter},
    city= {Beijing},
    country = {China}}

\begin{abstract}

It has been reported that the Large High Altitude Air Shower Observatory (LHAASO) observed very high energy photons from GRB 221009A, with the highest energy reaching 18~TeV. We find that observation of such high energy photons is quite nontrivial since extragalactic background light could absorb these photons severely and the flux is too weak to be observed. Therefore we discuss a potential mechanism for us to observe these photons, and suggest that Lorentz invariance violation induced threshold anomaly of the process \(\gamma\gamma\to e^-e^+\) provides a candidate to explain this phenomenon.

\end{abstract}

\begin{keyword}

Lorentz invariance violation, threshold anomaly, very high energy photons, gamma-ray burst 

\end{keyword}

\end{frontmatter}


On 9 October 2022, a special gamma-ray burst (GRB) was reported first by Fermi and Swift and is numbered as GRB 221009A~\cite{GBM1, GBM2, LAT1, LAT2, Swift1, Swift2, Swift3}. This burst, located at around RA = 288.282 and Dec = 19.495~\cite{LAT2}, is a long burst but with a very small redshift \(z=0.1505\)~\cite{Redshift1, Redshift} compared to most other long bursts. More remarkably, the extraordinary brightness of this burst immediately makes this burst a very interesting object for studying various aspects of physics and astronomy. The highest energy photon observed by Fermi-LAT reaches 99.3 GeV~\cite{LAT2}, while more strikingly, the Large High Altitude Air Shower Observatory (LHAASO) reported more than 5000 very high energy photon events with energies above 500~GeV, including photons with the energy up to 18~TeV, making these photons the most energetic GRB photons ever observed~\cite{LHAASO}. This unexpected observation of 18~TeV photons immediately brings a new question: can we observe photons with this energy if we consider only the standard knowledge of the Universe and particle physics? In the following we study the background light absorption of such photons in the Universe, show that we may need to invoke new mechanism to explain this observation. We also argue that Lorentz invariance violation~(LIV) induced threshold anomaly~\cite{KLUZNIAK1999117,Kifune:1999ex,Protheroe:2000hp,Mattingly2003, Jacobson2003,Li:2021pre2,LI20212254,LI2021a,lihao} provides a potential solution.

The Fermi Gamma-Ray Burst Monitor was triggered at 13:16:59.99 UT on 9 October 2022, and then located GRB 221009A~\cite{GBM1, GBM2}.  
LHAASO also joined in the observation of this GRB\@. Since LHAASO is designed for gamma-ray astronomy in the energy band between \(10^{11}\) and \(10^{15}~\text{eV}\)~\cite{Cao2010,DISCIASCIO2016166, Cao2019, Bai:2019khm}, it eventually reveals the very high energy property of this burst. The most intriguing result of LHAASO is that there are photons of about 18~TeV~\cite{LHAASO}. These photons can be used to help us understand the features of GRBs, and also we can utilize them to study the propagation of high energy photons in the Universe. According to the standard model of particle physics, two photons can annihilate with each other and produce an electron-positron pair: \(\gamma\gamma\to e^-e^+\). This annihilation process then prohibits high energy photons from propagating a long distance in the Universe since they can react with background photons such as those from cosmic microwave background~(CMB) and extragalactic background light~(EBL). Let \(E\) and \(\varepsilon_b\) be the energies of the high energy photon and the background photon respectively, then according to special relativity we can easily derive the threshold for this reaction to happen~\cite{LI2021a,lihao}:
\begin{equation}
    E\ge E_{\text{th}}=\frac{m_e^2}{\varepsilon_b},\label{thres}
\end{equation}
where \(m_e\) is the mass of electrons. For CMB photons, the mean energy is \(\varepsilon_b\simeq 6.35\times10^{-4}~\text{eV}\), resulting in a threshold energy \(E_{\text{th}}^\text{CMB}\simeq 411~\text{TeV}\)~\cite{LI2021a, lihao}, which indicates that CMB is almost transparent to the LHAASO photon of 18~TeV. However, for EBL photons the situation could be more complicated. As a rough estimate, we can take \(\varepsilon_b\simeq 10^{-3}~\text{eV}\) to \(1~\text{eV}\) according to the distribution of EBL, leading to the corresponding thresholds \(E_{\text{th}}\simeq 261~\text{GeV}\) to \(261~\text{TeV}\)~\cite{lihao}. As a result, EBL photons do matter to cause an attenuation of cosmic photon propagation in the analysis of the LHAASO photons of above 10~TeV~\cite{lihao}. In the following, after adopting one of the conceivable EBL models~\cite{Dominguez2011}, we discuss the indications 
of LHAASO very high energy photons from GRB 221009A in detail.

\begin{figure}
    \centering
    \includegraphics[scale=0.28]{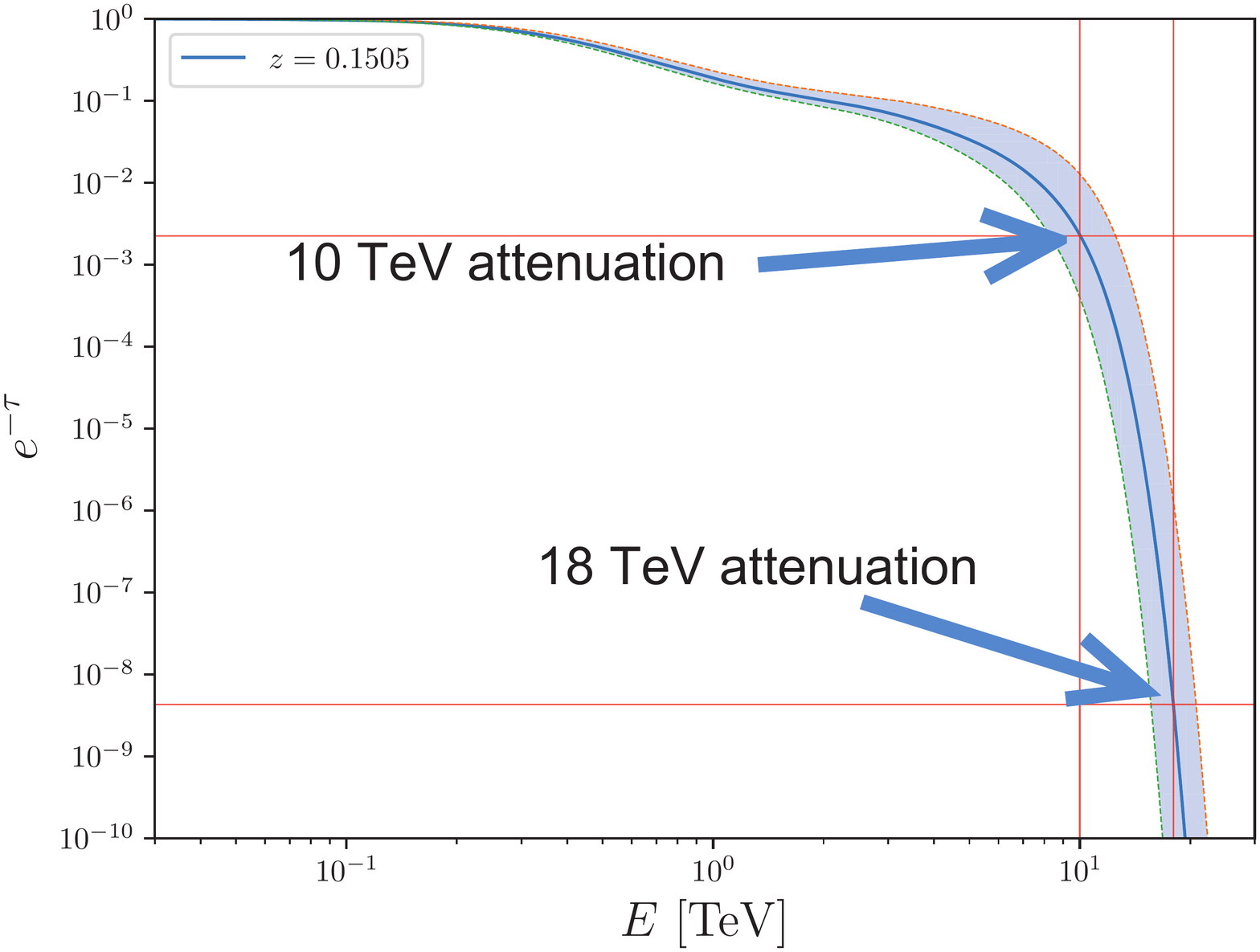}
    \caption{The EBL attenuation~(\(e^{-\tau}\)) of photons from GRB 221009A with redshift \(z=0.1505\). The model and data are taken from Ref.~\cite{Dominguez2011} and \url{http://side.iaa.es/EBL/}.\label{tau}}
\end{figure}

\begin{figure}
    \centering
    \includegraphics[scale=0.28]{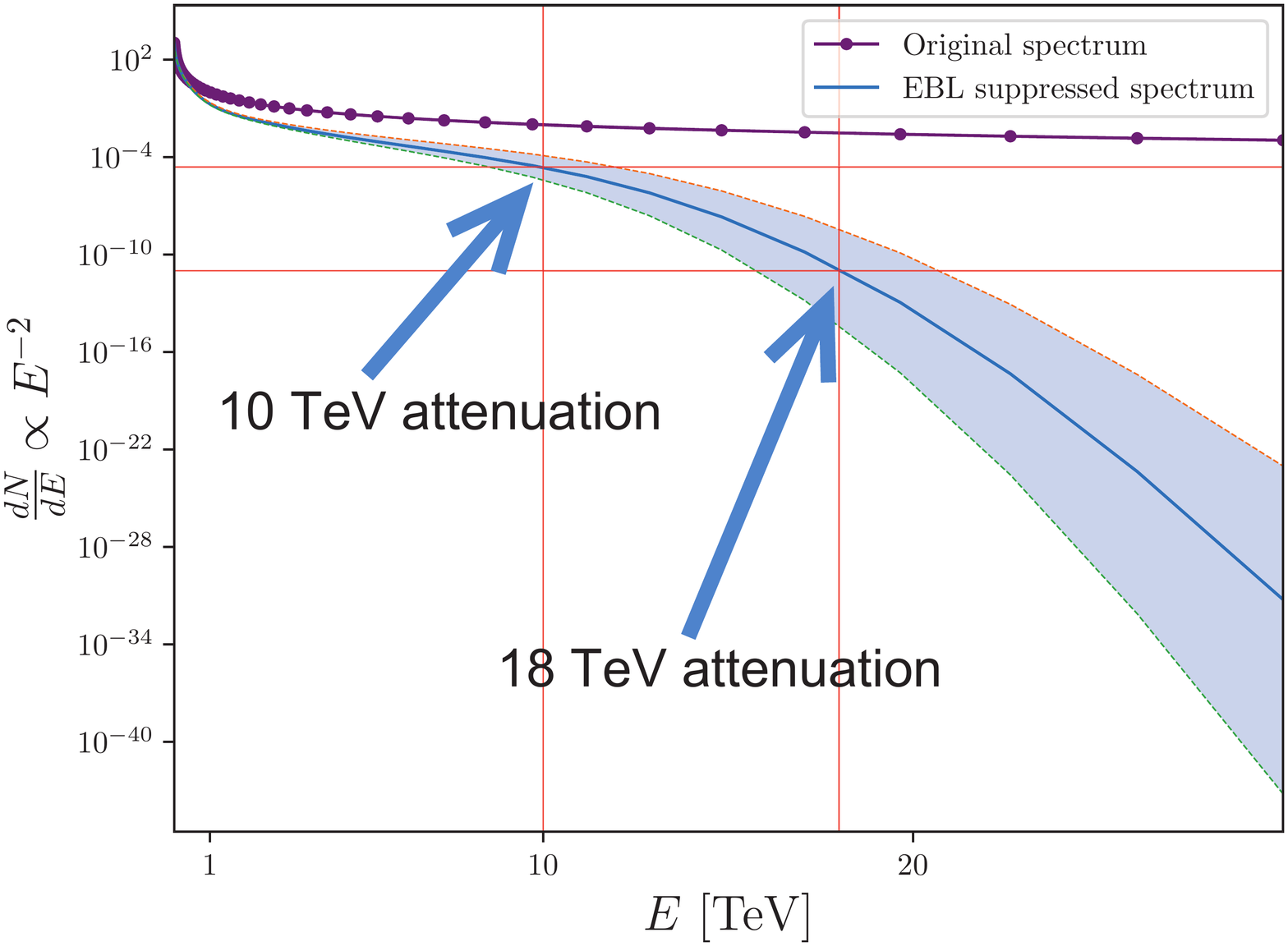}
    \caption{The original and EBL suppressed spectra of \(dN/dE\propto E^{-2}\).\label{spectrum}}
\end{figure}

To calculate the EBL attenuation of photons from GRB 221009A, we have to know the distribution of EBL photons. However it is out of the scope of this short paper to discuss too many details of EBL, we suggest the readers to consult Refs.~\cite{Dominguez2011,Gilmore2012,lihao}. Here we just adopt the model and result of~\cite{Dominguez2011,lihao}. According to~\cite{Dominguez2011,Gilmore2012}, we have to calculate the optical depth \(\tau(E_\text{obs}, z)\) where \(E_\text{obs}\) is the observed energy of the photon and \(z\) is the redshift. Then let us denote the intrinsic and observed fluxes as \(F_\text{int}\) and \(F_\text{obs}\) respectively, they obey the relation:
\begin{equation}
    F_\text{obs} = F_\text{int} \times e^{-\tau}. \label{flux}
\end{equation}
For \(z=0.1505\), we draw the \(E\)-\(e^{-\tau}\) plot in Fig.~\ref{tau}, where we also mark out the points for 10~TeV photons and 18~TeV photons for later discussions. We find that for 10~TeV photons, the flux is suppressed by about several hundred times, and it is still possible to observe these photons since the suppression is not too restrictive. However for 18~TeV photons the situation is totally different. We can infer from the picture that for 18~TeV photons the flux is suppressed by at least \(10^{-8}\), which almost forbids any observation of these photons. Even we consider the upper value of the EBL attenuation within uncertainties due to the explicit background photon distribution with energy dependence, a factor of about \(10^{-6}\) still makes the flux too weak to be detected. To illustrate the result better, assuming that the intrinsic flux can be approximated by
\begin{equation}
    \frac{dN}{dE}\propto E^{-2},
\end{equation}
we show the original spectrum and the suppressed one in Fig~\ref{spectrum}. We conclude that it is still possible to observe 10~TeV photons, but the observation of 18~TeV photons might be taken as a possibility for deviation from the standard physics. 

The EBL model picked for the above calculations has been compared with other EBL models in the original EBL model paper, see, Fig.~13 in Ref.~\cite{Dominguez2011}.
From which we see that the background EBL photons with larger energy (small $\lambda$) are more suppressed than other 
EBL models, so other models should predict smaller mean free paths than the EBL model we adopted. Therefore the EBL model including its uncertainties is more constrained than other EBL models but with larger mean free path. Choice of other model calculations would offer more strong conclusion that the observation of 18 TeV photons suggests the possibility for new physics.

As an explanation of the observation of 18~TeV very high energy photons by LHAASO with energies up to 18~TeV, we suggest that Lorentz invariance violation~\cite{Amelino-Camelia1997a, Amelino-Camelia1998, Mattingly2005, Amelino-Camelia2013, He:2022gyk} may be a potential candidate. As a consequence of LIV, the dispersion relation of photons may be expressed as
\begin{equation}
    \omega(k)\simeq k^2-\xi k^3,\label{dispersion}
\end{equation}
in which \(\xi>0\)~(\(\xi<0\)) corresponds to the subluminal (superluminal) case of Lorentz violating photons~\cite{He:2022gyk}. 
If we assume the standard energy-momentum conservation law and ignore the LIV effects of electrons and positrons\footnote{In fact, the LIV effect of electrons and positrons is found to be rather small from the analysis of LHAASO data~\cite{Li:2022ugz}.}, the threshold for the process \(\gamma\gamma\to e^-e^+\) is modified and we obtain the following picture describing the threshold anomaly for different values of \(\xi{}\)~\cite{LI2021a}:
\begin{description}
  \item[Case I] If \(\xi>\xi_c={16\varepsilon_b^3}/({27m_e^4})\), subluminal photons cannot be absorbed by background photons with energy \(\varepsilon_b\) through the process \(\gamma\gamma\to e^- e^+\). 
  \item[Case II] If \(0<\xi<\xi_c\), a subluminal photon can be absorbed only when its energy falls into a certain closed interval with the lower bound larger than \(E_\text{th}\). This means that there is an upper threshold and the \(\varepsilon_b\) background is again transparent to photons whose energies exceed this upper threshold. 
  \item[Case III] If \(\xi<0\), the threshold behavior resembles that of the case in special relativity, except that the modified threshold now is smaller than \(E_{\text{th}}\).
\end{description}
In fact, the threshold anomaly of \(\gamma\gamma\to e^- e^+\) in high energy cosmic photon propagation has been extensively studied in literature~\cite{Mattingly2003, Jacobson2003,Li:2021pre2,LI20212254}.
There is an optional constraint on \(\xi{}\) from other phenomenological studies, and the result reads \(\xi^{-1}\geq 3.6\times 10^{17}~\text{GeV}\)~\cite{Shao2010f,Zhang2015,Xu2016a,Xu2016,Amelino-Camelia:2016ohi,Amelino-Camelia:2017zva,Xu2018,Liu2018,Li2020,Zhu2021a,Chen2021}, with possible range falling into the first case and as a result EBL might be transparent to these 18~TeV photons\footnote{The suggested numerical value of \(\xi^{-1}\) 
just serves as an example to demonstrate our viewpoint. Actually the permissible range of \(\xi{}\) is quite wide, only requiring that \(E_\text{LV}:=\xi^{-1}\lesssim 1.15\times 10^{23}~\text{GeV}\simeq 10^{4}\times E_\text{Planck}\), consistent with available constraints for subluminal photons.}. Therefore the observation of 18~TeV photons might be a consequence of the transparency of the background light to high energy photons in the Universe. Indeed even if the second case is the realistic one, one might still explain the observation of 18~TeV photons from GRB 221009A as long as \(\xi{}\) takes suitable values.

In summary, we find that the LHAASO observation of very high energy photons with energies up to 18~TeV from GRB 221009A is an extraordinary result.  Within the framework of standard model and general relativity, EBL should absorb these photons with energies above 10~TeV severely and any observatories should be hard to observe photons with energies of or above 18~TeV. 
Here we suggest that LIV induced threshold anomaly~\cite{KLUZNIAK1999117,Kifune:1999ex,Protheroe:2000hp,Mattingly2003, Jacobson2003,Li:2021pre2,LI20212254,LI2021a,lihao} can serve as a potential mechanism for this observation and future observation or detection of other very high energy photons from GRBs can testify our suggestion. After all, besides LIV, there are still many other potential explanations such as the EBL model might be modified with background photon distribution at larger energy more constrained, or the photons are just from a near source that is coincident with GRB 221009A both spatially and temporally~(although this is very unlikely to happen), 
or the 18~TeV photons are produced through conventional electromagnetic cascade by the electron-positron pair produced by a photon with energy much higher than 18~TeV during the propagation to the Earth,
or the uncertainty in the photon energy reconstruction is too big so that the real photon energy might be well smaller than 18~TeV, or the existence of the axion-photon convention to convent photons into axion-like particles near the source and then to re-convert these axion-like particles into photons near the detector~\cite{axion-like}. With more data in the future, we should be able to constrain some options and arrive at a more concrete conclusion.

\section*{Declaration of competing interest\label{declaration}}

\declaration{}

\section*{Acknowledgements\label{acknowledgements}}

This work is supported by \fund.


\end{document}